\documentclass[preprint]{elsarticle}

\usepackage{amssymb}
\usepackage{amsbsy}
\usepackage{amsmath}
\usepackage{color}
\usepackage{pdflscape}
\usepackage{afterpage}
\usepackage{capt-of}
\usepackage{amsfonts}
\usepackage{graphicx}
\usepackage{epstopdf}
\usepackage{algorithmic}
\usepackage{hyperref}
\usepackage{cleveref}
\usepackage{comment}
\usepackage{epstopdf}
\usepackage[algo2e,ruled,linesnumbered]{algorithm2e}
\usepackage{makecell}
\usepackage{subfigure}
\usepackage{float}
\usepackage{pdfpages}
\usepackage{mymacros}

\ifpdf
  \DeclareGraphicsExtensions{.eps,.pdf,.png,.jpg}
\else
  \DeclareGraphicsExtensions{.eps}
\fi

\def\n {\tilde{n}}
\renewcommand{\req}[1]{Eq.~\ref{eq:#1}}

\newcommand{\Uin}[2]{f_{#1}^{#2}}

\makeatletter
\newcommand{\mathleft}{\@fleqntrue\@mathmargin2em}
\newcommand{\mathcenter}{\@fleqnfalse}
\makeatother

\journal{}

\begin{document}

\begin{frontmatter}



\title{A scalable weakly-synchronous algorithm for solving partial differential
equations}


\author[label1]{Konduri Aditya\corref{cor1}\fnref{fnt1}}
\ead{konduriadi@iisc.ac.in}
\fntext[fnt1]{Current affiliation: Department of Computational and Data Sciences, Indian Institute of Science, Bangalore, India}
\author[label2]{Tobias Gysi}
\ead{tobias.gysi@inf.ethz.ch}
\author[label2]{Grzegorz Kwasniewski}
\ead{grzegkwas@gmail.com}
\author[label2]{Torsten Hoefler}
\ead{htor@inf.ethz.ch}
\author[label3]{Diego A.\ Donzis}
\ead{donzis@tamu.edu}
\author[label1]{Jacqueline H.\ Chen}
\ead{jhchen@sandia.gov}
\cortext[cor1]{Corresponding author.}
\address[label1]{Combustion Research Facility, Sandia National Laboratories, Livermore, CA 94550, United States}
\address[label2]{Department of Computer Science, ETH Zurich, 8092, Zuirch, Swizterland}
\address[label3]{Department of Aerospace Engineering, Texas A\&M University, College Station, TX 77843, United States}

\begin{abstract}
Synchronization overheads pose a major challenge as applications advance
towards extreme scales. In current large-scale algorithms, synchronization
as well as data communication delay the parallel computations at each time
step in a time-dependent partial differential equation (PDE) solver. This
creates a new scaling wall when moving towards exascale.
We present a weakly-synchronous algorithm based on novel
asynchrony-tolerant (AT) finite-difference schemes that relax
synchronization at a mathematical level. 
We utilize remote memory access programming schemes 
that have been shown to provide significant
speedup on modern supercomputers, to efficiently implement communications
suitable for AT schemes, and compare to two-sided communications that are state-of-practice.
We present results from simulations of Burgers' equation as a model of 
multi-scale strongly non-linear dynamical systems. Our algorithm
demonstrate excellent scalability of the new AT schemes for large-scale
computing, with a speedup of up to $3.3$x in communication time and $2.19$x in total runtime.
We expect that such schemes can form the basis for exascale PDE
algorithms.
\end{abstract}


\end{frontmatter}



\section{Introduction}
Numerical simulations are an important tool in the pursuit of understanding critical problems in science and engineering.
Many natural phenomena and engineered systems are described using highly non-linear partial differential equations (PDEs).
At the conditions of practical interest, non-linearity in the equations results in a multi-scale phenomena which demands massive computations at extreme levels of parallelism.
Current state-of-the-art simulations are routinely being carried out on hundreds of thousands of processing elements (PEs) \cite{DJ2013,DASY2014,MC2016,LMM2013}.
It has been observed at large scale that data synchronizations between PEs and communications result in poor scalability of applications.
On future exascale machines, which will consist of millions of PEs, the
scalability of applications is expected to be further compromised.
There are several efforts at various levels of the stack (hardware, software
and algorithms) to relax the data synchronization and avoid
communication~\cite{doi:10.1137/090760969,Grigori:2008:CAG:1413370.1413400,Donfack:2014:DBS:2672598.2672900}.

\begin{figure}[h]
  \centering
  \includegraphics[trim=6cm 8cm 5cm 8cm,clip,width=0.75\linewidth]{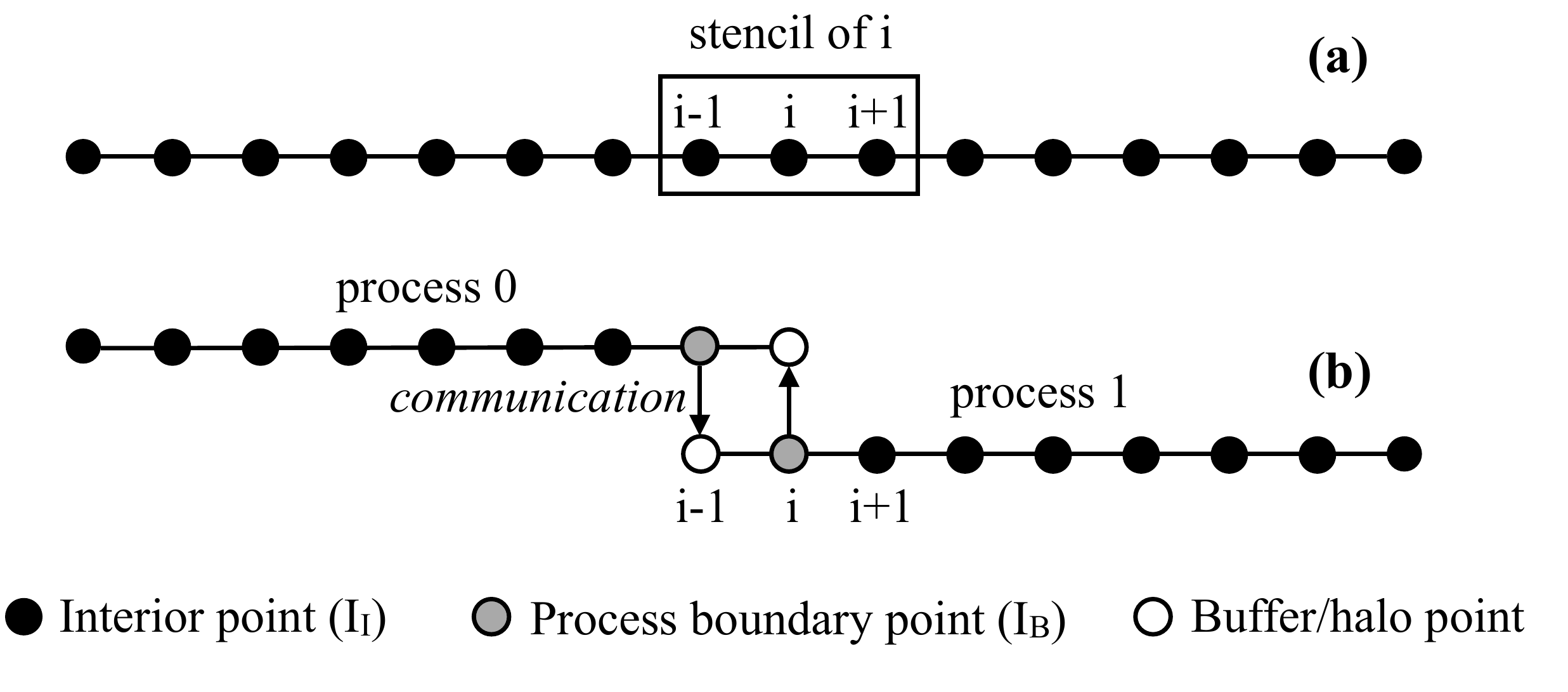}
  \caption{Schematic of discretized one-dimensional domain decomposed into two processes ($P=2$).}
  \label{fig:grid-s}
\end{figure}

In time-dependent PDE solvers, the governing equations are approximated with spatial and temporal schemes which result in algebraic equations.
These equations are solved at each grid point in a discrete computational
domain (e.g., see \autoref{fig:grid-s}~(a)), for given initial and boundary
conditions, to obtain the solution of a function at a time instant or level.
This process is repeated in a time-marching or iterative manner to arrive at a time evolved solution.
Typically, computations at a generic point $i$ depend on the value of the function at its neighbors.
The extent of the neighborhood is determined by the spatial schemes, which is referred to as a stencil.
To parallelize the computations in a solver, the set of grid points in a domain 
is decomposed into $P$ subsets that are then distributed to $P$ compute
processes, as illustrated in \autoref{fig:grid-s}~(b).
If not all the stencil points for a grid point $i$ are in the same subset (i.e., $i$ is a process boundary point), then the function values of the missing points have to be communicated to the owner process of point $i$, before the computations are advanced.
This communication is often facilitated through the so called buffer or halo points, which can be understood as a local cache of the remote values.
These buffers can also be seen as the special case of a one-time producer-consumer queue: at each iteration producers communicate new values into the buffer and consumers wait until the data becomes available.
We call PDE algorithms that rely on synchronized buffer data to advance computations from a time level $n$ to $n+1$ as {\em synchronous algorithms}.

In simple synchronous algorithms, both receivers have to wait for
senders and senders may have to wait for receivers. The latter
synchronization can be relaxed either implicitly with message
buffering~\cite{mpi-3}, which is feasible only for small messages, or
explicitly with multi-version variables that expose the buffer as a
producer-consumer queue with multiple
entries~\cite{dotsenko2007expressiveness}. Yet, the receiver
still has to wait for the sender to guarantee \emph{consistent} data in
the buffer. Due to the iterative nature of PDE computations, this
synchronization prevents two processes from running more than one iteration
apart in real-time---a very stringent requirement for large-scale
systems where delays are normal and propagate
quickly~\cite{hoefler-noise-sim}.

With increasing degree of parallelism, the size of the point set per
process shrinks and, relatively more stencil points are at remote processes. Thus, the
communication and synchronization overhead grows steadily while the
iteration compute time per process shrinks.
On future exascale systems, it can be expected that the communication
time is equivalent or higher than the computation time, thus, the smallest
process delays will lead to high synchronization overheads. In addition,
future large-scale systems are expected to exhibit a higher performance
variability than today's systems~\cite{Shalf:2010:ECT:1964238.1964240},
further aggravating the overheads due to process synchronization.

This work presents a method to relax the synchronization further such that the
receiver does not wait for the sender and computations at process boundary points are performed with potentially outdated values.
The concept and mathematical feasibility of such an asynchronous computational
approach for PDEs has been studied \cite{DA2014,AD2017}, and is summarized in
\autoref{sec:concept}.
New asynchrony-tolerant (AT) finite-difference schemes
\cite{AD2017} will be used to design a novel and accurate weakly-synchronous algorithm for
PDE solvers, which can proceed with computations of $L$ time level advancements 
in an asynchronous fashion (data at halo points can be up to $L$ iterations
apart).
The performance of the algorithm will be compared with commonly used synchronous algorithm
to demonstrate
the scalability of weakly-synchronous (WS) algorithm at large scales. Two different communication models,
one-sided remote memory access and two-sided message passing, are used to evaluate the effect of different parallel programming models.

The rest of the paper is organized as follows. The mathematical background of asynchronous computing for PDEs is described in \autoref{sec:concept}.
Simulation details of a benchmark problem are presented in \autoref{sec:sim}. The parallel programming models to implement data communication are described in \autoref{sec:progmodels}.
\cref{sec:implementation} illustrates the synchronous and weakly-synchronous algorithms and their implementation. Results on numerical accuracy and computational performance are reported in \autoref{sec:results}. Related work and conclusions are discussed in \cref{sec:related,sec:conclusions}, respectively.

\section{Mathematical Background}
\label{sec:concept}
In a PDE solver, the basic requirement of data synchronization between PEs is imposed by the numerical method, and in particular, computation of the spatial derivatives.
When this is relaxed at a mathematical level, other synchronizations in the computing algorithm can also be relaxed.
To understand this in a mathematical perspective, consider the computations at 
the grid point $i$ in \autoref{fig:grid-s}~(b), which assumes a three point 
stencil.
Let $f$ be a function whose solution is advanced from a time level $n$ to $n+1$ in an iteration.
As the owner process of $i-1$ is different from $i$, the value of $f$ at $i-1$ is communicated into the halo point at each iteration. We refer to these communications as halo exchanges. 
In synchronous algorithms, computations at $i$ cannot proceed unless $f$ at the buffer point $i-1$ is at time level $n$. 
As mentioned earlier, this is ensured by explicit synchronization.
The spatial derivatives in synchronous algorithms are computed with uniform time level of the function at all the points in a stencil. For explicit-in-time schemes, the function will be at the time level $n$.
We refer to such computations as {\emph {synchronous computations}}.

In the asynchronous computing approach \cite{KD2012,DA2014}, values of $f$ are communicated in halo 
exchanges at each time iteration, however the synchronization is not imposed.
This means that the value of $f$ in the halo points may not be at time level $n$. Depending on the speed of communication, the time level can be from one of the $n$, $n-1$, $n-2$, $\dots$ levels.
Let $\n= n-\k{}$ denote the latest time level of $f$ available at a halo point, where $\k{}$ indicates the delay in terms of number of levels.
If $\n = n$ or $\k{}=0$, then the time level of $f$ at a halo point is synchronous. Otherwise, if $\n<n$ or $\k{}>0$, then $f$ is at an asynchronous or delayed time level.
As the delay cannot be indefinite, we restrict the maximum allowable delay to $L$ time levels, which means $\n\in \{n,n-1,\dots,n-L+1\}$. If the delay at a halo point exceeds $L$, then a synchronization of messages has to be imposed to reduce the delay value to be less than $L$.
For this reason, we refer to the algorithm for the asynchronous computing approach as a {\emph{weakly-synchronous algorithm}}.

Considering a non-zero probability of having an asynchronous value of $f$ at halo points, let us examine the nature of computations carried out at grid points in the domain. 
For interior points (see \autoref{fig:grid-s}~(b)), all the points in their 
stencil are computed in the same process, and the function values will be at a 
time level $n$. Hence, computations at these points are synchronous.
On the other hand, some of the points in the stencil of a process boundary point are halo points, and the time level $\n$ of $f$ at the halo points can be synchronous or asynchronous. The computations at process boundary points are synchronous if $\n=n$ (i.e., $f$ is synchronous). Otherwise, the time level of $f$ at different points in the stencil is non-uniform and  we refer to these computations at process boundary points as {\emph{asynchronous computations}}. Clearly, the nature of computations at process boundary points depends on how fast the messages are delivered. 

An analysis based on the finite-difference method \cite{DA2014} showed that 
numerical properties of standard schemes, when computations are performed in an
asynchronous fashion, will not only depend on numerical parameters like the 
grid resolution ($\dx$) and time step ($\dt$), but also on simulation 
parameters like the number of processes ($P$) used in a simulation and the 
characteristics of communication or the delay ($\k{}$). 
It has also been demonstrated that while standard schemes are able to maintain stability as well as consistency, their accuracy is significantly affected.
The reduced accuracy is attributed to terms in the truncation error of schemes 
that appear due to asynchrony. 
In a subsequent study \cite{AD2017}, new asynchrony-tolerant (AT) schemes with arbitrary order of accuracy have been derived.
These schemes use a wider stencil to eliminate the low order asynchrony terms in the truncation error and result in a higher order accuracy.  
The wider stencil of AT schemes, relative to standard schemes, can be obtained either by adding more points in space or by using multiple time levels of the function at each stencil point. The former leads to larger message sizes and the latter increases the memory requirements of a process.
The AT schemes used in this paper are shown below. 
The expressions for the schemes evaluated at a grid point $i$ and time level $n$ are, for a delay on the left side of the stencil,
\bea
\left.\doe{f}{x}\right|_i^n &=& \frac{\Uin{i+1}{n}-(\k{}+1)\Uin{i-1}{n-\k{}}+\k{}\Uin{i-1}{n-\k{}-1}}{2\dx} + \ord\left(\dx^2\right) \nonumber\\
\left.\frac{\pdd f}{\pd x^2}\right|_i^n &=& \frac{\Uin{i+1}{n}-2\Uin{i}{n}+(\k{}+1)\Uin{i-1}{n-\k{}}-\k{}\Uin{i-1}{n-\k{}-1}}{\dx^2} \nonumber\\
 &&  + \ord\left({\k{}(\k{}+1)}{\dt^2}/{\dx^2},\dx^2\right). \label{eq:AT1}
\eea
And, for a delay on the right side of the stencil,
\bea
\left.\doe{f}{x}\right|_i^n &=& \frac{(\k{}+1)\Uin{i+1}{n-\k{}}-\k{}\Uin{i+1}{n-\k{}-1}-\Uin{i-1}{n}}{2\dx} + \ord\left(\dx^2\right) \nonumber\\
\left.\frac{\pdd f}{\pd x^2}\right|_i^n &=& \frac{(\k{}+1)\Uin{i+1}{n-\k{}}-\k{}\Uin{i+1}{n-\k{}-1}-2\Uin{i}{n}+\Uin{i-1}{n}}{\dx^2} \nonumber\\
&  & +  \ord\left({\k{}(\k{}+1)}{\dt^2}/{\dx^2},\dx^2\right). \label{eq:AT2}
\eea
Here, the second term on the right hand side of each expression represents the truncation error with leading order terms shown in the parenthesis. These schemes are second order accurate in space when the time step in the simulations varies as $\dt\sim\dx^2$.
It can be shown \cite{AD2017} that, to leading order, the average error ($\la E\ra$) due to asynchronous computations for these schemes is 
\be 
\la E\ra \sim \frac{P}{N} \dx^{2} \left( \kmom{2} + \kmom{} \right),
\label{eq:err_scaling}
\ee
where, $N$ is the number of grid points in the domain, and $\kmom{2}$ and $\kmom{}$ are moments of the delay $\k{}$ measured during the course of the simulation.
This equation shows that the error scales linearly with an increase in the number of processes, and goes up with an increase in the degree of asynchrony. 
However, these affects are diminished by the term $\dx^2$, which provides a control parameter to obtain solutions of desired accuracy.


\section{Simulation Details}\label{sec:sim}
To demonstrate the implementation of the weakly-synchronous algorithm, we choose a problem of high relevance in practical applications.
This section describes the problem definition and the governing PDEs, the numerical method, and the choice of programming model for communications between the processes.

\subsection{Problem Definition}\label{sec:problem}
The viscous Burgers' equation provides a rich model to understand the multi-scale phenomena in fluid flows. 
We simulate Burgers' turbulence in a three-dimensional domain as a benchmark problem in this paper.
In addition to the equations governing the velocity field, we also solve scalar advection-diffusion equations to replicate the computational complexities in many fluid flow solvers.
Scalars are often used in simulations of turbulence to understand its mixing properties \cite{DASY2014}.
In reacting flow solvers \cite{Aditya2018}, which are used to simulate combustion processes, scalars are transported representing different species in a chemical mechanism.
The governing equations of the problem are 
\bea
  \frac{\pd u_{i}}{\pd t} + u_j \doe{u_i}{x_{j}}
 & = & \nu \frac{\pdd u_i}{\pd x_j^2} \hspace{0.5cm} \text{ for $j=1,2,3$  and} \nonumber\\
  \frac{\pd \phi_{k}}{\pd t} + u_j\doe{\phi_k}{x_{j}}
 & = & \alpha_k \frac{\pdd \phi_k}{\pd x_j^2},
\label{eq:BS}
\eea
where $u_i$ ($i=1,2,3$) represents the velocity component in the $i$th direction and $\nu$ is the kinematic viscosity of the fluid. 
$\phi_k$ and $\alpha_k$ are the concentration and diffusivity of the $k$th scalar, respectively.
The equations are solved in a periodic domain with equal sides.
Initial conditions are derived from a multi-scale sinusoidal function.

\subsection{Numerical Method}
\label{sec:method}

\begin{figure}
      \centering
      \includegraphics[width=0.5\linewidth]{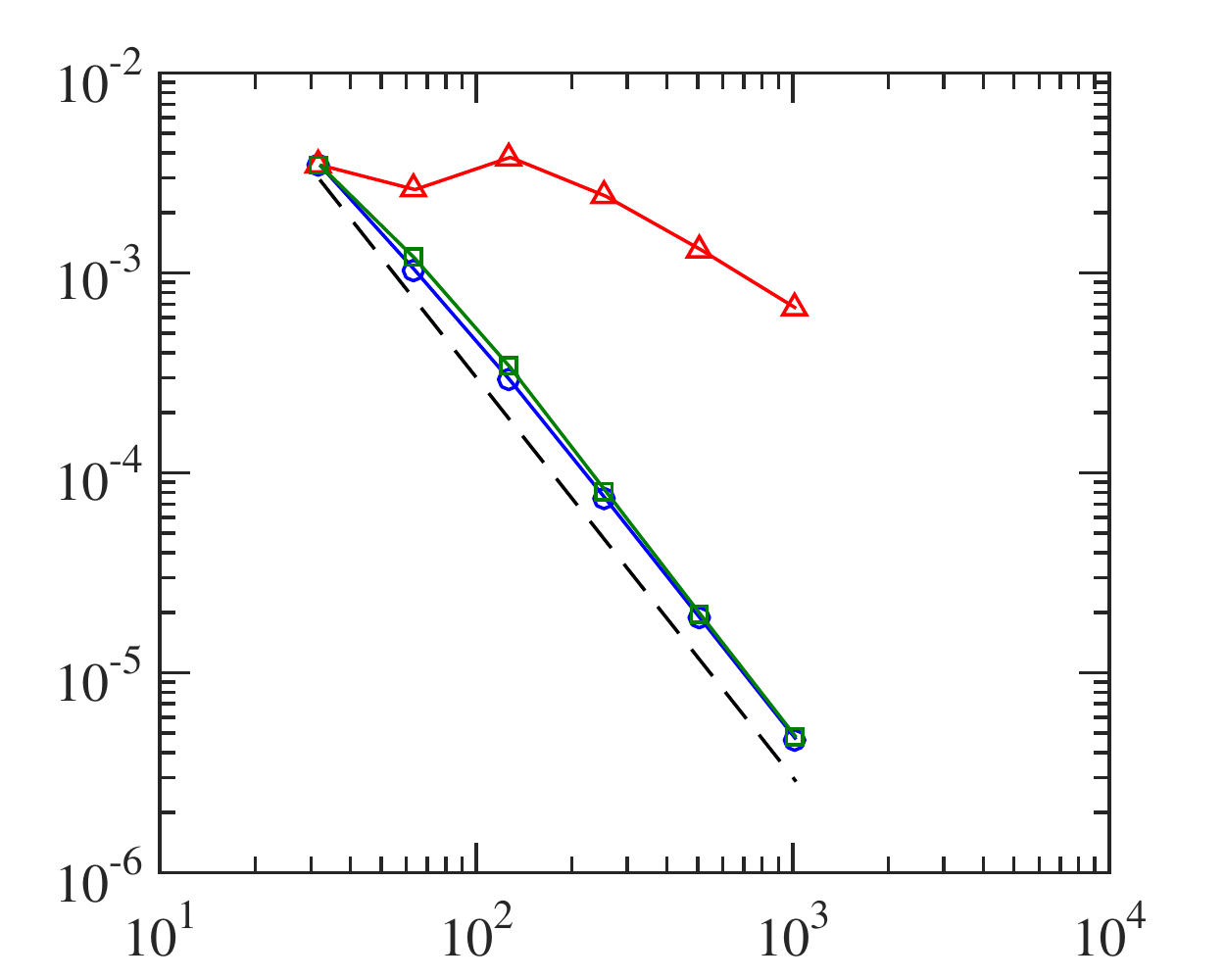}
      \bp
      \put(-140,-5){Number of grid points}
      \put(-193,45){\rotatebox{90}{Average error}}
      \put(-120,65){$-2$}
      \ep
      \caption{Graph of average error convergence with increase in grid points. 
      In the graph: (blue-circle line) configuration with synchronous algorithm and 
      standard schemes, (red-triangle line) configuration with weakly-synchronous 
      algorithm and standard schemes, (green-square line) configuration with 
      weakly-synchronous algorithm which uses standard schemes for synchronous 
      computations and AT schemes for weakly-synchronous computations, 
      (black line) a reference line with slope equal to $-2$. Simulations are 
      done with $P=8$.}
      \label{fig:accuracy}
\end{figure}

The PDEs in \req{BS} are solved using the finite difference method.
To aid the description of numerical schemes for synchronous and weakly-synchronous algorithms, we define different subsets of grid points in a process. 
Let $I$ denote the set of points where the solution of $u_i$ and $\phi_k$ is computed.
The set $I$ is further divided into $\Ii$ and $\Ib$.
$\Ii$ is the set of interior points whose computations do not depend on communication between processes.
The points near process boundaries belong to set $\Ib$, and their computations depend on communications.

Time integration is performed with the Euler scheme that is first order accurate in time and with a stability condition based on viscous/diffusive process ($\dt \sim \dx^2$), which results in second order accuracy in space.
The spatial derivatives at a grid point $i\in\Ii$ are computed using standard second order accurate central difference schemes.
At process boundary points, i.e., $i\in \Ib$, the stencil computations are 
asynchronous when there is a delay in communication. 
We use AT schemes in \reqs{AT1}{AT2} to compute the spatial derivatives.
Note that for $\k{}=0$, these AT schemes reduce to the standard central difference schemes used at interior points. 
This helps in homogenizing the error in the domain and lower the overall error in the solution \cite{AD2017}.

Before proceeding to the next section, the affect of asynchrony on the accuracy of standard central difference schemes and the role of AT schemes used for computations in the set $\Ib$ are demonstrated. 
Numerical simulations of a one-dimensional linear advection-diffusion equation are used for this purpose. The numerical schemes and initial and boundary conditions are the same as the one described above for the Burgers' turbulence problem. The advection-diffusion equation has a simple analytical solution for this setup, which is used to compute the exact error in the computed solution and to provide a verification for the accuracy of algorithms.
The simulations are performed in three different configurations.
The first configuration computes with a synchronous algorithm using standard schemes. The second configuration computes with the weakly-synchronous algorithm using standard central difference schemes for points in sets $\Ii$ and $\Ib$. In the third configuration, the weakly-synchronous algorithm with standard schemes for points in $\Ii$ and AT schemes for points in $\Ib$ are used.
\autoref{fig:accuracy} shows how the error in the solution varies with an increase 
in the number of grid points. For the first configuration (blue-circle line), which 
uses the synchronous algorithm, the error decreases with a slope equal $-2$. This is 
expected since the schemes used are second order accurate.
When the standard schemes are used in a weakly-synchronous algorithm 
(red-triangle line), due to the asynchronous computations at process boundaries 
the accuracy reduces drastically. However, this loss in accuracy of the 
weakly-synchronous algorithm is recovered by using AT schemes at $\Ib$, as 
indicated by the green-square line in the figure which has a slope equal to 
$-2$.
Aditya et al. \cite{AD2017} present a more detailed analysis for arbitrary order of accuracy.

\section{Parallel Programming Models}
\label{sec:progmodels}

The asynchronous computing approach imposes three requirements that parallel programming models should facilitate. First, they should enable communication without explicit synchronization at each time step. Second, they should store data from multiple time levels. 
And third, they should have knowledge of time level associated with each message. In terms of performance, the communication speed between process, which depends on the parallel programming model, not only effects the scalability, but also the numerical accuracy of the solution (e.g., as shown in \req{err_scaling}). Hence, it is important to choose a fast, low-overhead parallel programming model that fulfills the above mentioned requirements in developing the weakly-synchronous algorithm.

\subsection{Two-sided MPI}

Two-sided MPI is the dominant parallel programming model in 
high-performance computing. To move data between the processes running on a 
distributed memory machine, the source and target processes call 
the MPI library to send and receive the data via the network. The 
popularity of two-sided MPI may be attributed to its performance and ease of 
use when developing bulk synchronous applications that alternate between 
computation and communication. However, every data transfer using two-sided MPI entails target interactions, such as the receive buffer address resolution, 
that prevent an efficient implementation of weak synchronization. 
The non-blocking interface suggests the possibility of overlapping 
communication and asynchronous computation. Yet, the limited hardware support 
necessitates periodic calls to the MPI library to guarantee asynchronous 
progress, which affects complexity and performance of a possible 
implementation~\cite{asyncprogress}. We use the non-blocking interface
to implement baseline variants of our communication algorithms.


\subsection{Notified Remote Memory Access}
\label{sec:notified}

One-sided MPI exploits the RDMA (Remote 
Memory Direct Access) capabilities of modern network interconnects 
to guarantee asynchronous progress.
Using one-sided MPI, only the source process calls the MPI library and provides all 
information necessary to move the data directly to the target memory. This 
direct access enables hardware acceleration and avoids target interactions, 
such as the receive buffer address resolution. Moreover, the source and target 
processes are able to progress independently, which is key for an efficient 
implementation of weak synchronization. 

Like most RMA (Remote Memory Access) implementations, one-sided MPI explicitly 
differentiates data movement and synchronization. This separation enhances the 
expressiveness and flexibility of the programming model at the cost of 
additional synchronization overheads. To avoid this performance penalty, the 
foMPI-NA~\cite{notified-access} programming model introduces notified RMA.
With foMPI-NA, the source process attaches a notification to the RMA. After 
completion of the data transfer, the network interconnect pushes the 
notification to a queue in the target memory. To synchronize, the target 
process queries the queue for incoming notifications. This mechanism enables 
efficient one-way synchronization using existing MPI design concepts.
The \textsf{MPI\_Win\_allocate} method allocates memory segments 
called windows that provide hardware supported RMA. The processes call the 
method collectively and use the window to address the remote memory. The 
\textsf{MPI\_Put\_notify} method writes to remote memory and after completion 
notifies the target with the given tag. The method uses the window and 
displacement parameters to address the remote memory of the target process. The 
\textsf{MPI\_Win\_flush\_all} method waits for the remote completion of pending accesses on the window. The \textsf{MPI\_Notify\_init} method implements notification 
matching based on persistent requests. The method parameters specify after how many notifications the matching completes. The matching filters 
notifications according to their tag and source process information. 
The \textsf{MPI\_Start} method resets the persistent requests and starts the 
notification matching. The \textsf{MPI\_Test} and \textsf{MPI\_Wait} methods 
test and wait for the completion of the notification matching. 

We implement the main variant of our communication algorithms with foMPI-NA since the 
programming model best fits the requirements of weak 
synchronization. The RMA semantics enables transferring data without target 
interaction and the notification queue enables querying the synchronization 
status without source interaction. As a result, the source and target processes 
can progress independently enabling maximum asynchrony.

\section{Implementation}
\label{sec:implementation}
\begin{figure*}
            \includegraphics[width=\textwidth]{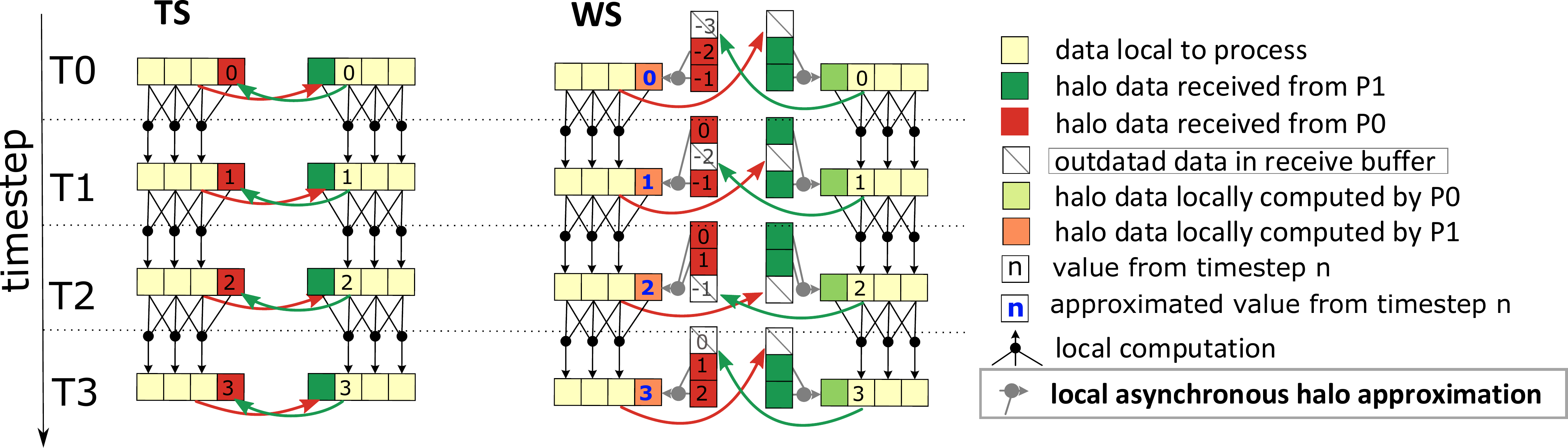}
            \caption{
                The TS algorithm synchronizes the processes after every time step,
                while the WS algorithm avoids synchronization if the process
                asynchrony stays within given bounds.
            }\label{fig:algorithms}
\end{figure*}

At scale, communication costs in PDE solvers are likely to dominate the overall 
runtime. This can be clearly observed in strong scaling a problem, which 
reduces the computational effort per process and significantly increases the 
number of communications. Also, the communication costs are often affected by 
system noise, which increase the costs further. Hence, recent advances in 
solver algorithms have a paradigm shift away from optimizing computations 
towards more efficient communication algorithms. These issues are addressed in 
the asynchronous computing approach, which naturally overlaps computations and 
communications, hiding the communication costs and reducing the impact of noise. In our weakly-synchronous algorithm, we exploit the benefits of 
the mathematical framework by designing an implementation with minimal compute 
and synchronization overheads.

We evaluate performance and scalability of the weakly-synchronous (WS) algorithm 
compared to the traditional synchronous (TS) algorithm. \autoref{fig:algorithms} 
explains the halo exchange communication performed by the two algorithms that 
otherwise share the time loop which besides the communication implements 
the stencil computations to advance the solution to the next time step.
The TS algorithm always exchanges the halo points of the current time step, 
which synchronizes the execution of the neighboring processes.
%
%
The WS algorithm always sends the halo points of the current time step and 
immediately advances the solution using the latest available halo points. The 
algorithm employs circular buffers to store the halo points of multiple 
time levels and updates the boundary points using the AT schemes 
discussed in \autoref{sec:method}. This approach avoids synchronization 
at the cost of few additional computations at comparable numerical accuracy. 

\begin{table}
	\centering
      \begin{tabular}{ll}
            \Xhline{2\arrayrulewidth}
            \textbf{name} & \textbf{description} \\
            \Xhline{2\arrayrulewidth}
            \textsf{N} & number of neighbors to communicate with \\
            \textsf{T} & number of entries in the circular buffers \\
            \textsf{L} & maximal allowable delay \\
            \textsf{sbuf/rbuf} & send/receive buffers \\
            \textsf{sbnd/rbnd} & bounds of the halo/boundary zones \\ 
            \textsf{roff} & offset of the receive buffers in the target window \\
            \textsf{nbr} & identifiers of the neighbor processes\\
            \textsf{rqst} & notification matching requests \\
            \textsf{data} & data array for local computations \\
            $t_s$ & send time step of the outgoing data \\
            $t_r$ & receive time step of the last incoming data \\
            $t_c$ & time step constraint (synchronize if $t_c > t_r$) \\
            $h_s$ & send head pointing to the buffer entry of $t_s$ \\
            $h_r$ & receive head pointing to the buffer entry of $t_r$ \\
            \Xhline{2\arrayrulewidth}
      \end{tabular}
      \caption{Main variables of our implementation}
      \label{tab:vars}
\end{table}

We implement the two algorithms with the one-sided foMPI-NA programming model (NA) discussed in \autoref{sec:notified} and for comparison provide variants based on the non-blocking interface of two-sided MPI (NB).
We allocate one send-receive buffer pair per neighbor to communicate the data along all six faces of the cuboidal sub-domains. When running the WS algorithm, we allocate additional receive buffers that store the halo points of the last $L$ time levels and use tags to distinguish incoming notifications and messages from the different time levels.
%
%
All notified access implementations place the send and receive buffers at different offsets within one window, move the data with notified RMA, and synchronize the execution using one notification per direction and time step.
All two-sided implementations pre-post non-blocking receives for the next time levels, move the data using non-blocking sends, and wait for the completion of the receive requests to synchronize the execution.
Pre-posting the receives is the only algorithmic difference when switching from one-sided to two-sided. The synchronous (TS-NB) algorithm pre-posts the receives for the next time level while the weakly-synchronous (TS-NB) algorithm pre-posts receives for maximal allowable delay \textsf{L} time levels.
Apart from this difference, the two-sided variants are equal to the notified access based variants of the weakly-synchronous (WS-NA) and synchronous (TS-NA) algorithm next discussed in detail.




\subsection{Traditional Synchronous Algorithm (TS-NA)}
\label{sec:sync}
The TS algorithm exchanges the halo points of the current time level 
after every time step and synchronizes computation and communication.
The algorithm overlaps the communications to the neighbor process
but waits for their completion before proceeding with the next time step.
Although this blocking design prevents any overlap of computation and communication,
the TS algorithm is widely used in today's PDE solvers. 
This ubiquity originates from the simplicity and maturity of the algorithm 
that enables high compute and communication throughput.

\begin{algorithm2e} 
    \SetKwArray{Nbr}{nbr}
    \SetKwArray{Rqst}{rqst}
    \SetKwArray{Sbuf}{sbuf}
    \SetKwArray{Rbuf}{rbuf}
    \SetKwArray{Roff}{roff}
    \SetKwArray{Sbnd}{sbnd}
    \SetKwArray{Rbnd}{rbnd}
    \SetKwArray{Data}{data}
    \SetKwData{NEI}{N}
    \SetKwArray{Win}{win}
    \SetKwFunction{Put}{MPI\_Put\_notify}
    \SetKwFunction{Flush}{MPI\_Win\_flush\_all}
    \SetKwFunction{Start}{MPI\_Start}
    \SetKwFunction{Test}{MPI\_Test}
    \SetKwFunction{Wait}{MPI\_Wait}
    \SetKwFunction{Pack}{pack}
    \SetKwFunction{Unpack}{unpack}
    
    \tcp{send boundary points}
    \For{$i \leftarrow 1$ \KwTo \NEI}{                          \label{sa-send0}
          \Sbuf{:,$i$} $\leftarrow$ \Data{\Sbnd{$i$}}\;   
          \Put{\Sbuf{:,$i$},\Nbr{$i$},\Win,\Roff{$i$}}\;
    }                                                                             \label{sa-send1}
    \Flush{\Win}\;                                                          \label{sa-flush}
    \tcp{receive halo points}
    \For{$i \leftarrow 1$ \KwTo \NEI}{                          \label{sa-recv0}
          \Start{\Rqst{$i$}}\; 
          \Wait{\Rqst{$i$}}\;
          \Data{\Rbnd{$i$}} $\leftarrow$ \Rbuf{:,$i$}\;   
    }                                                                             \label{sa-recv1}
    \caption{Synchronous halo exchange}
    \label{alg:sync}
\end{algorithm2e}

\autoref{alg:sync} shows the synchronous halo exchange (TS-NA). 
The algorithm supports a configurable number of neighbors \textsf{N}. 
The arrays \textsf{sbuf} and \textsf{rbuf} contain one send and receive buffer per neighbor. 
The arrays \textsf{sbnd} and \textsf{rbnd} store the index ranges that locate the boundary 
and halo points in the \textsf{data} array. The \textsf{nbr} array stores the 
process identifiers that in combination with the window provide access to the 
neighbor memories. The \textsf{roff} array stores the receive buffer offsets 
per neighbor. The offsets are relative with respect to the window. The 
\textsf{rqst} array stores one request per neighbor that matches one 
notification sent by the particular neighbor. We initialize the persistent requests
at program startup and call the start method to reset them during the program execution.

On lines~\ref{sa-send0}--\ref{sa-flush}, 
we send the boundary points to all neighboring processes. We first 
pack the boundary points to the send buffer and then call the notified put 
method. We address the remote memory by combining process identifier, window, 
and receive buffer offset. After the send phase, we call the flush method to 
guarantee the completion of the sends.

On lines~\ref{sa-recv0}--\ref{sa-recv1}, we receive the halo points of all 
neighboring processes. We call the start and wait methods to match the incoming
notifications. Once the data is available, we unpack the receive buffer to the halo points.

\begin{algorithm2e} 
    \SetKwArray{Nbr}{nbr}
    \SetKwArray{Rqst}{rqst}
    \SetKwArray{Sbuf}{sbuf}
    \SetKwArray{Rbuf}{rbuf}
    \SetKwArray{Roff}{roff}
    \SetKwArray{Sbnd}{sbnd}
    \SetKwArray{Rbnd}{rbnd}
    \SetKwArray{Data}{data}
    \SetKwArray{TR}{$t_r$}
    \SetKwData{TS}{$t_s$}
    \SetKwData{TC}{$t_c$}
    \SetKwData{LEV}{T}
    \SetKwData{NEI}{N}
    \SetKwData{DLY}{L}
    \SetKwArray{Win}{win}
    \SetKwFunction{Put}{MPI\_Put\_notify}
    \SetKwFunction{Flush}{MPI\_Win\_flush\_all}
    \SetKwFunction{Start}{MPI\_Start}
    \SetKwFunction{Test}{MPI\_Test}
    \SetKwFunction{Wait}{MPI\_Wait}
    \SetKwFunction{Max}{max}
    \SetKwFunction{Min}{min}
    \SetKwFunction{Extra}{approximate}
    \tcp{communicate boundary}
    \Flush{\Win}\;                                                          \label{wa-send0}
    $t_s \leftarrow t_s + 1$\tcp*{update send time}
    $h_s \leftarrow \TS \mod{\LEV}$\tcp*{map to circular buffer offset}
    \For{$i \leftarrow 1$ \KwTo \NEI}{
          \Sbuf{:,$i$} $\leftarrow$ \Data{\Sbnd{$i$}}\;   
          $d \leftarrow \Roff{$i$} + h_s \times |\Sbuf{:,i}|$\; 
          \Put{\Sbuf{:,$i$},\Nbr{$i$},\Win,$d$,$h_s$}\;
    }                                                                             \label{wa-send1}
    \tcp{consume notifications}
    \eIf{$\TS \leq \LEV$}{                                            \label{wa-dlay0}
          $\TC \leftarrow \TS$\tcp*{enforce synchrony during init}
          \Flush{\Win}\;
    }{
          $\TC \leftarrow \TS - \DLY$\tcp*{enable asynchrony}
    }                                                                             \label{wa-dlay1}
    \For{$i \leftarrow 1$ \KwTo \NEI}{                          \label{wa-recv0}
          $f \leftarrow true$\tcp*{test flag}
          \While{$f \land \TS > \TR{i}$ }{                      \label{wa-test0}
                $h_r \leftarrow (\TR{i}+1) \mod{\LEV}$\tcp*{next request offset}
                \Start{\Rqst{$h_r$,$i$}}\; 
                \Test{\Rqst{$h_r$,$i$},$f$}\;
                \If{$f$}{
                      $\TR{i} \leftarrow \TR{i} + 1$\tcp*{increment receive time}
                }
          }                                                                       \label{wa-test1}
          \While{$t_c > \TR{i}$ }{                                    \label{wa-wait0}
                $h_r \leftarrow (\TR{i}+1) \mod{\LEV}$\tcp*{next request offset}
                \Start{\Rqst{$h_r$,$i$}}\; 
                \Wait{\Rqst{$h_r$,$i$}}\;
                $\TR{i} \leftarrow \TR{i} + 1$\tcp*{increment receive time}
          }                                                                                   \label{wa-wait1}
    }                                                                                   \label{wa-recv1}
    \tcp{approximate the halo points}         
    \For{$i \leftarrow 1$ \KwTo \NEI}{                          \label{wa-extra0}
          \Data{\Rbnd{i}} $\leftarrow$ \Extra{\Rbuf{:,:,$i$},\TS,\TR{i}}\;  
    }                                                                             \label{wa-extra1}
    \caption{Weakly-synchronous halo exchange}
    \label{alg:async} 
\end{algorithm2e}

To further tune our implementations, we hoist the pack and unpack logic to 
separate OpenMP parallel loops that enable threading. 
 However, we do not 
overlap the communication with the computation on an inner domain. This 
optimization divides the innermost loops in three separate loops which impairs 
the strong scaling performance. For example, the memory access patterns 
become more complex and the vectorization overheads increase.




\subsection{Weakly-synchronous Algorithm (WS-NA)}

The WS algorithm sends the halo points of the current time level after every time step. 
But the algorithm does not wait for the incoming data of the 
current time step, and instead employs AT schemes that update the boundary 
points using past time levels. This non-blocking design enables efficient 
overlap of computation and communication and avoids synchronization to the 
point. The algorithm synchronizes the communications only when the 
halo point delay exceeds the maximal allowable 
value. This is required to ensure sufficient numerical accuracy.  

The implementation of the WS algorithm allocates circular buffers to store 
the halo points of the most recent time levels. The neighbors write to 
the circular buffers using the displacement associated with the current time 
step and tag the notified access with the corresponding circular buffer 
offsets. Moreover, the AT schemes apply synchronization dependent stencils at 
the sub-domain boundary, which complicates the code and potentially impairs 
performance. We therefore split the AT schemes into the synchronization 
dependent halo point approximation and the stencil computation. The 
approximation phase updates the halo points using the most recent circular 
buffer entries, while the stencil computation remains unchanged with respect 
to the TS algorithm. This approach maintains the memory access patterns of the 
TS algorithm except for the halo point approximation that may consume multiple 
time levels.

\autoref{alg:async} shows the weakly-synchronous halo 
exchange (WS-NA). The algorithm uses the same arrays as the
synchronous algorithm but extends the receive buffer array \textsf{rbuf}
and the notification requests array \textsf{rqst} to circular buffers
for all \textsf{T} time levels.
The variables $h_s$ and $h_r$ point to the send and receive 
head of the circular buffers. To compute the head pointers, the variable 
$t_s$ and the array $t_r$ keep track of the current send and receive time 
steps. 
The variable $t_c$ limits the difference of the send and receive time 
steps to the maximal allowable approximation delay \textsf{L}. The variable $d$ defines the  displacement in the target window.
We set the circular buffer size \textsf{T} to twice the maximal allowable delay \textsf{L} plus the number of time levels accessed by the stencil. 

On lines~\ref{wa-send0}--\ref{wa-send1}, 
we send the boundary points to all neighboring processes. 
We start the communication by calling the flush method to complete the sends 
of the previous time step. This late flush is possible since the WS algorithm 
cannot deadlock due to missing data from the current time step. 
We next increment the send time step and assign its value modulo 
array size to the send head. The displacement in the target window then 
corresponds to the product of the send head and the send buffer size plus the 
receive buffer array offset. To communicate the data, we pack the boundary 
points to the send buffer and call the notified put method with the tag set to 
the send head.  

On lines~\ref{wa-dlay0}--\ref{wa-dlay1}, 
we define the synchronization target of the halo point approximation.
At the beginning of the program execution we synchronize the execution after every time step to initialize the circular buffers. We thus set the time step constraint to the send time step and call the flush method to complete the data transfers. After the initialization, we relax the synchronization by subtracting the maximal allowable approximation delay from the send time step.

On lines~\ref{wa-recv0}--\ref{wa-recv1},
we update the receive time step for all neighboring processes. 
We first increment the receive time step by matching incoming notifications one-by-one in  
chronological order. To query the next notification, we increment the receive 
time step modulo array size and wait for the associated notification request. 
The receive time step update happens in two phases.

On lines~\ref{wa-test0}--\ref{wa-test1}, 
we perform the non-blocking update of the receive time step using 
the test method that returns true in case the notification 
matching request completed and false otherwise. The non-blocking update either 
advances the receive time step up to the first pending notification request or 
stops at the send time step.

On lines~\ref{wa-wait0}--\ref{wa-wait1},
we perform the blocking update of the receive time step using 
the wait method that blocks until the notification matching 
request completed. The blocking update ensures that the receive time step does 
not fall behind the time step constraint. 

On lines~\ref{wa-extra0}--\ref{wa-extra1}, 
we approximate the halo points using the most recent receive buffer entries.

\begin{figure}
	 \centering
      \includegraphics[width=0.75\columnwidth]{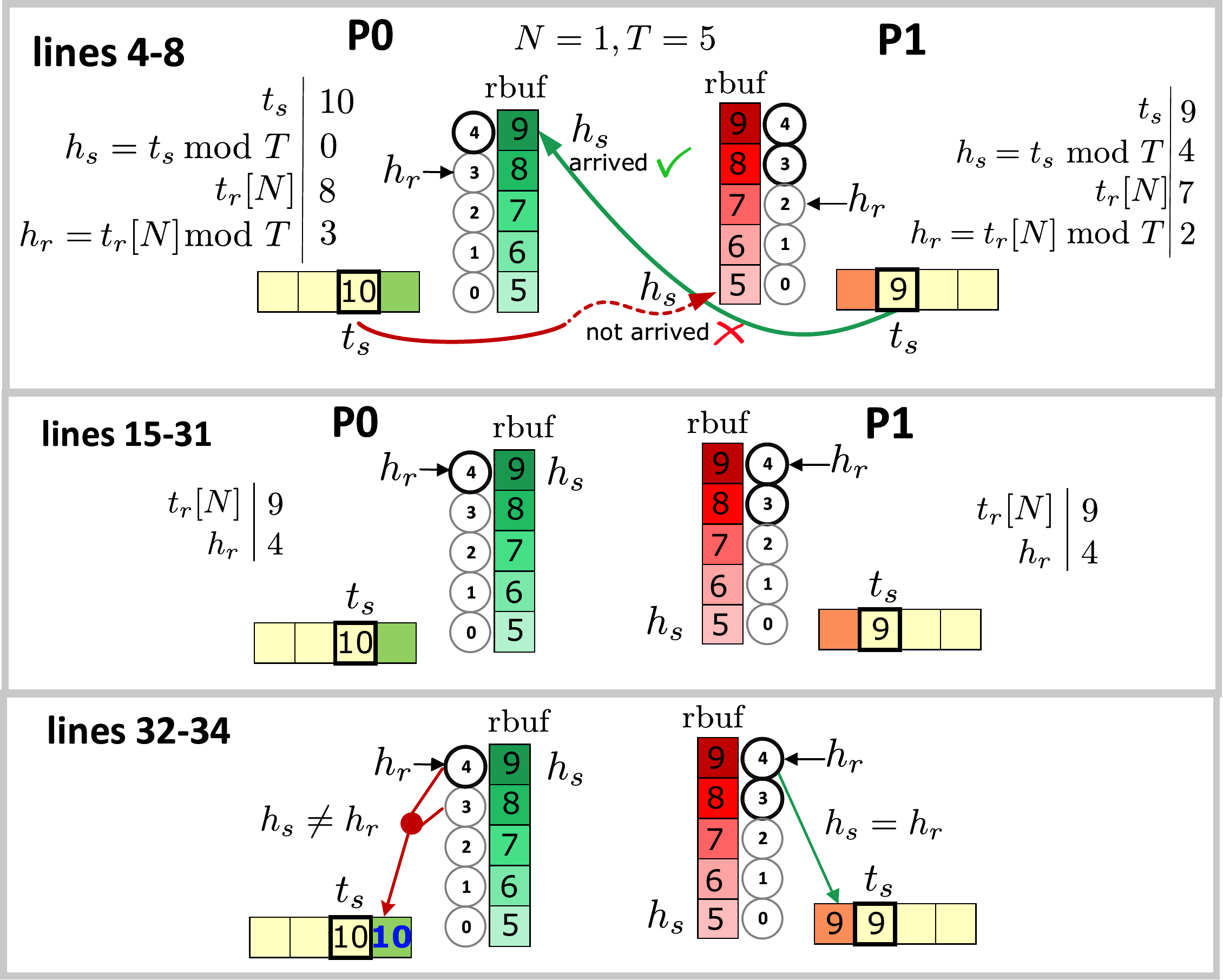}
      \caption{
            The processes P0 and P1 communicate using the weakly-synchronous halo 
            exchange algorithm: the write of P1 completes while P0 
            just starts writing (lines 4-8), both P0 and P1 consume the 
            notifications and advance the receive heads (lines 15-31), and finally 
            P0 and P1 update the halo points using the latest available data 
            (lines 32-34).}
      \label{fig:aexchange}
\end{figure}

\autoref{fig:aexchange} illustrates halo exchange 
communication. The processes P0 and P1 execute the time steps 10 and 9 which are 
composed of three main phases:
1) the processes write the halo points of the current time level to the 
corresponding entry in the circular buffer of the neighbor process, 2) the 
processes P0 and P1 consume the notifications and advance the receive heads 
accordingly, and 3) the processes P0 and P1 update the halo points using the 
data at the receive head. The process P1 copies the up-to-date data, while the 
process P0 approximates the halo points using two time levels.

We again tune the performance by hoisting the pack and approximation logic to OpenMP parallel loops. We also vary the start direction of the send and synchronization loops in round-robin fashion to avoid directional synchronization bias. 
%


\section{Results}\label{sec:results}
In this section, we evaluate performance of the proposed weakly-synchronous (WS) algorithm against the traditional-synchronous (TS) algorithm.
Before presenting the results on 
numerical accuracy and computational performance, we will briefly describe the 
simulation parameters and methodology used to obtain the results.

\subsection{Setup and Methodology}

To benchmark the algorithms, we developed a mini-application that solves the 
governing equations. The code is written in Fortran90 and parallelized 
with a hybrid OpenMP-MPI model. The experimental evaluation was performed on 
the GPU partition of the Piz Daint supercomputer at the Swiss National 
Supercomputing Center CSCS. The Cray XC50 connects 4400 nodes each with 
1 Xeon E5-2690 v3 CPU and 1 Tesla P100 GPU using an Aries network 
interconnect with Dragonfly topology. All experiments launch 1~MPI process 
and 12~OpenMP threads per node, which fully utilizes the 12-core 
CPU while the GPU remains idle. We compile our code with gfortran 6.2 and
use the foMPI-NA-0.2.4 (NA) and mpich2-1.2.1 (NB) libraries to implement
the communication.

For the evaluation of numerical
accuracy, simulations at different grid resolution are computed to a same end
time of the solution. 
Unlike TS algorithm, which result in a unique solution for a given set of numerical and simulation parameters, the accuracy in the case of the WS algorithm is expected to vary with repetitions of the same experiment. This is due to the variability in the function time delay at halo points. Hence, we repeat each experiment 15 times in different allocations. A constant maximum allowable delay of $L=10$ has been used in the simulations of the WS algorithm.

For the computational performance experiments, we first execute 30 warm-up time steps and then 1000 time steps in a steady state for the measurements. The warm-up steps are required to populate the circular buffers of the WS algorithm. Again, a constant maximum allowable delay of $L=10$ is used in the case of WS algorithm simulations.
As compute and communication times are commonly affected by system noise, we repeat each experiment 15 times in different allocations, as performed in the numerical accuracy experiments. 
We report median values and visualize the lower and upper quartiles using error bars. 

\subsection{Numerical Accuracy}
In \autoref{fig:accuracy} and \autoref{sec:method}, we have demonstrated the formal accuracy of the numerical method by showing the convergence of error in the computed solution obtained from the simulations of the linear advection-diffusion equation.
Due to non-linearity in the advection term, the equations governing Burgers' turbulence do not possess a simple analytical solution to compute the error.
We, therefore, evaluate the numerical accuracy of the WS algorithm by comparing its solutions against the TS algorithm solution.

\begin{figure}
     \centering
     \subfigure{\includegraphics[trim=1cm 1cm 1cm 1cm,clip,width=0.4\linewidth]{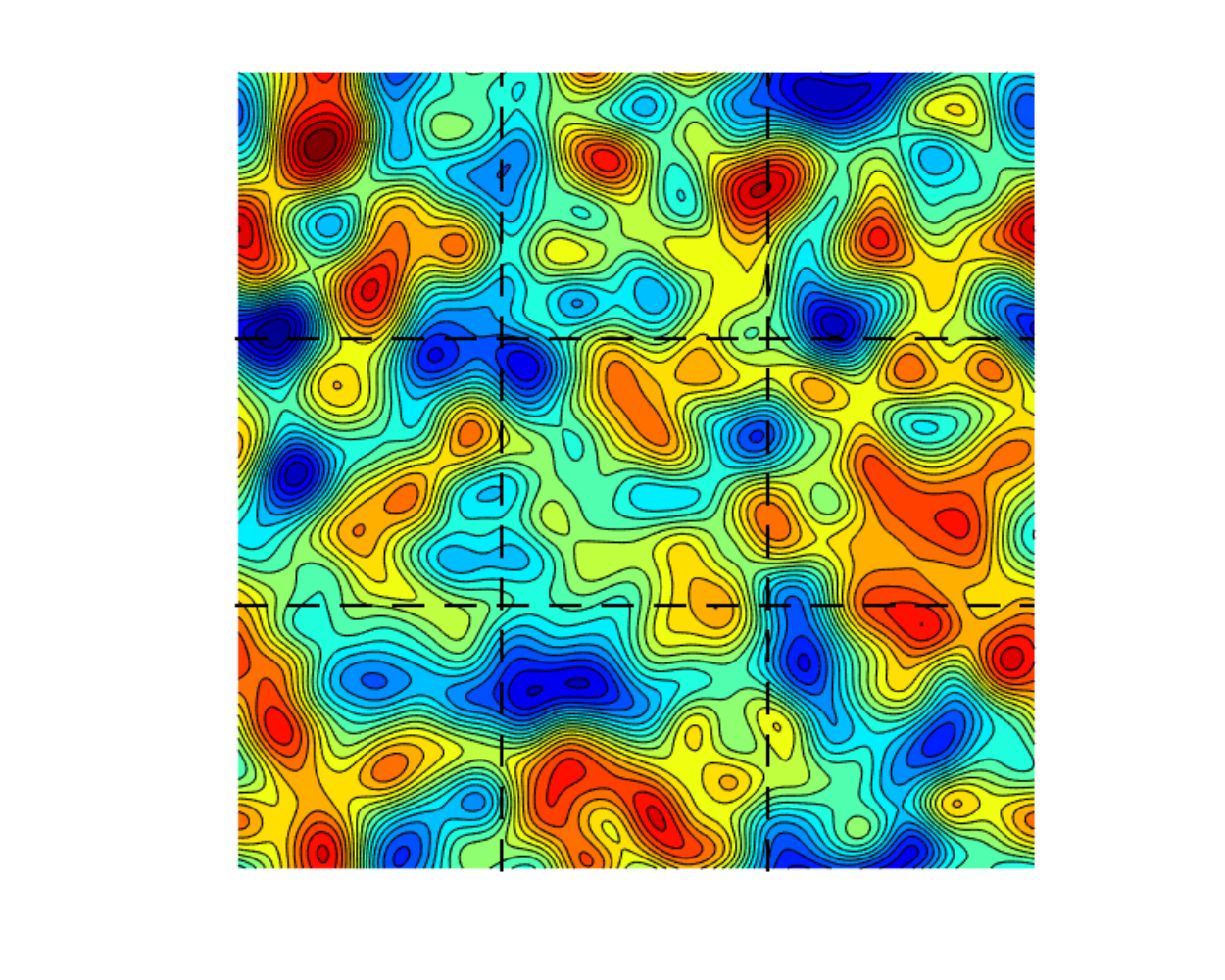}}
     \subfigure{\includegraphics[trim=1cm 1cm 1cm 1cm,clip,width=0.4\linewidth]{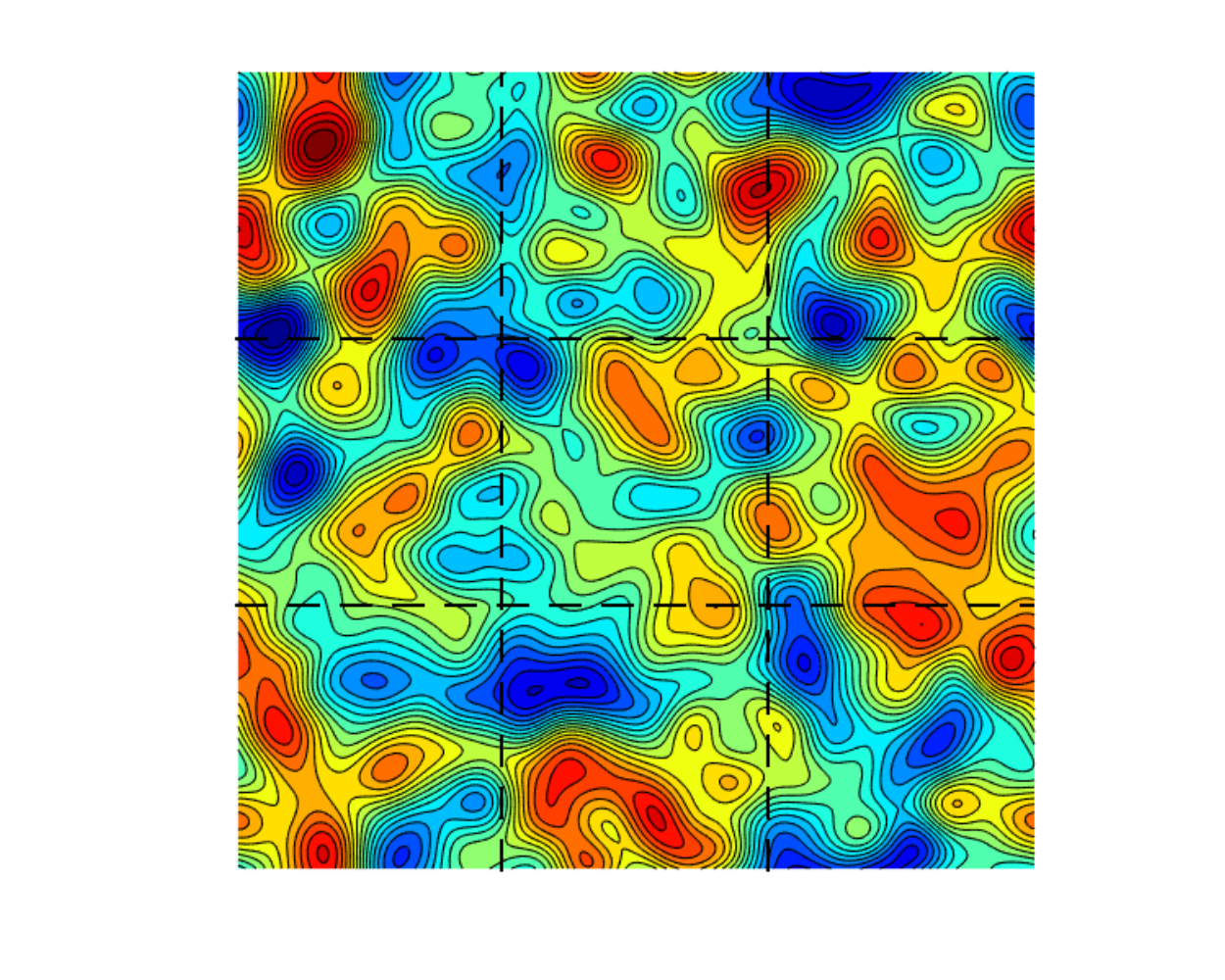}}
     \put(-298,100){\bf (a)}
     \put(-147,100){\bf (b)}
     \caption{Instantaneous conoturs of velocity component $u_1$ obtained 
from simulations using the (a) TS and (b) WS algorithms. The dashed black 
lines 
represent the process boundaries associated with the domain decomposition.}
     \label{fig:contours}
\end{figure}

The asynchronous computations can introduce large errors at the process boundaries due to delayed function values in the stencil. However, the use of AT schemes at the process boundaries improves the numerical properties and restricts the error to remain within the order of accuracy.
To examine the affects of asynchrony, instantaneous contours of the velocity component $u_1$ in a 2D plane obtained from the two algorithms using notified access communications are shown in \autoref{fig:contours}. The grid resolution is $\ngp{N}=\ngp{120}$ and 3 MPI processes along each dimension have been used in the simulations. The multi-scale nature of the solution is evident from the contours.
In comparison with the TS algorithm solution, the WS algorithm solution does not exhibit any noticeable aberrations either at the process boundaries, where the error due to asynchrony is introduced, or in the interior regions of the sub-domains, where the asynchrony error is expected to propagate with time.  

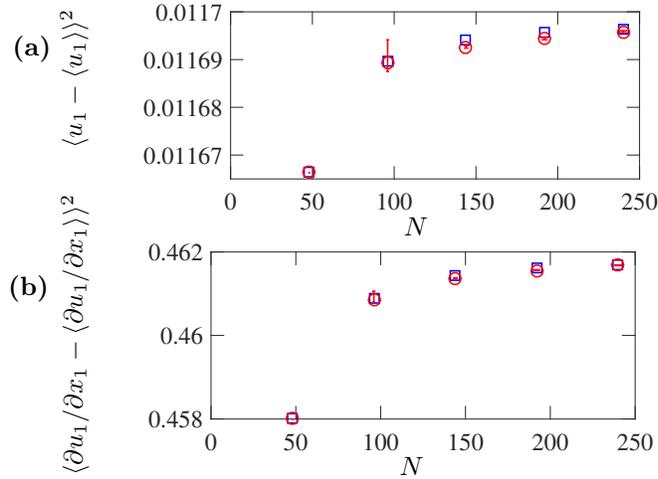
\begin{figure}
      \centering
      \includegraphics[width=0.6\linewidth]{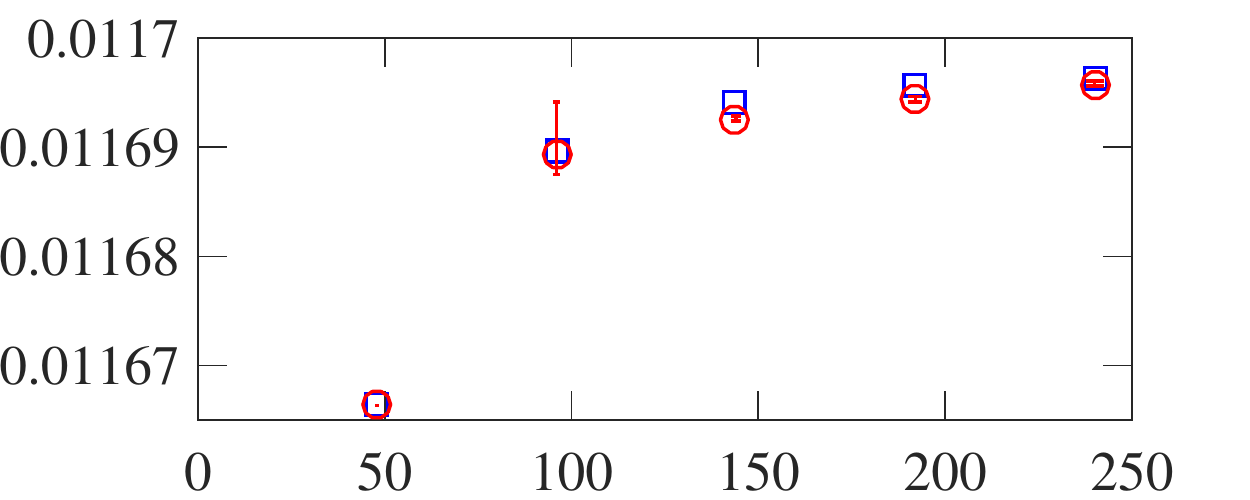}
      \begin{picture}(0,0)
      \put(-112,-8){$N$}
      \put(-242,25){\rotatebox{90}{$\xave{u_1-\xave{u_1}}^2$}}
      \put(-260,60){\bf (a)}
      \end{picture}

      \vspace{0.25cm}
      \includegraphics[width=0.6\linewidth]{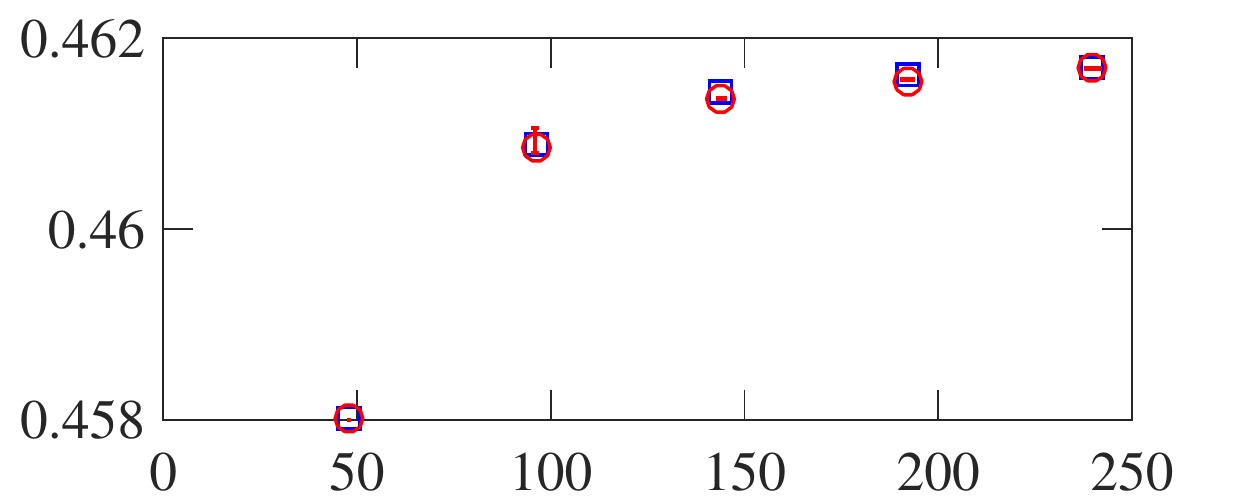}
      \begin{picture}(0,0)
      \put(-112,-8){$N$}
      \put(-240,-8){\rotatebox{90}{$\xave{\pd u_1/\pd x_1 -\xave{\pd u_1/\pd 
      x_1}}^2$}}
      \put(-260,60){\bf (b)}
      \end{picture}
      \caption{Variation of second central moment of (a) velocity component $u_1$ 
      and (b) velocity-gradients component $\pd u_1/\pd x_1$ with grid count $N$. 
      Blue-squares and red-circles are obtained from TS and WS 
      algorithms, respectively. The error bars in red-circles represent the variability in values among different runs. } 
      \label{fig:at-conv}
\end{figure}

A grid convergence study is performed, which is a common practice in computational fluid dynamics, is performed to qualitatively assess the numerical accuracy.
In such a study, values of key quantities of interest are computed at different grid counts.
For a stable and an accurate numerical method, the error in the numerical solution decreases with an increase in the grid count. This results in the values of key quantities to asymptote to a constant value.
In our experiments, we compute the convergence of moments of velocity and velocity gradients to study the effects of asynchrony on the accuracy of the WS algorithm.
\autoref{fig:at-conv} (a) and (b) show the grid convergence of the second moment of velocity component $u_1$ and its longitudinal gradient $\pd u_1/\pd x_1$, respectively.
In the graphs, $N$ is the number of grid points in a given direction.
With an increase in $N$ from $48$ to $240$, the number of hardware nodes have been increased from $1$ to $125$ in the simulations, which corresponds to a weak scaling of the problem with $\ngp{48}$ grid points per process.
We observe that the moments in both the graphs converge to an asymptotic value for both TS (blue squares) as well as WS (red circles) algorithms.
At the highest grid count the difference between the values from the two algorithms is less than $0.005\%$.
The confidence intervals are shown for the moments from WS algorithm to show the effect of varying asynchrony among different simulation runs. The intervals show a very narrow spread at higher grid counts, which demonstrates that for a grid converged solution the results are insensitive to the degree of asynchrony.
In these experiments, we have used a small set of simulation parameters ($L=10$, $P$ from $1$ to $125$) to evaluate the accuracy. 
It is very likely that the error in the solution would change for other simulation parameters. With an increase in $L$, the AT schemes at process boundaries will keep the error bounded as long as the overall numerical method remains stable \cite{DA2014,AD2017}.
This study demonstrates that the WS algorithm provides an accurate solution 
for relaxed synchronization between processes.

\subsection{Computational Performance}
\begin{figure}
      \centering
      \includegraphics[width=0.45\linewidth]{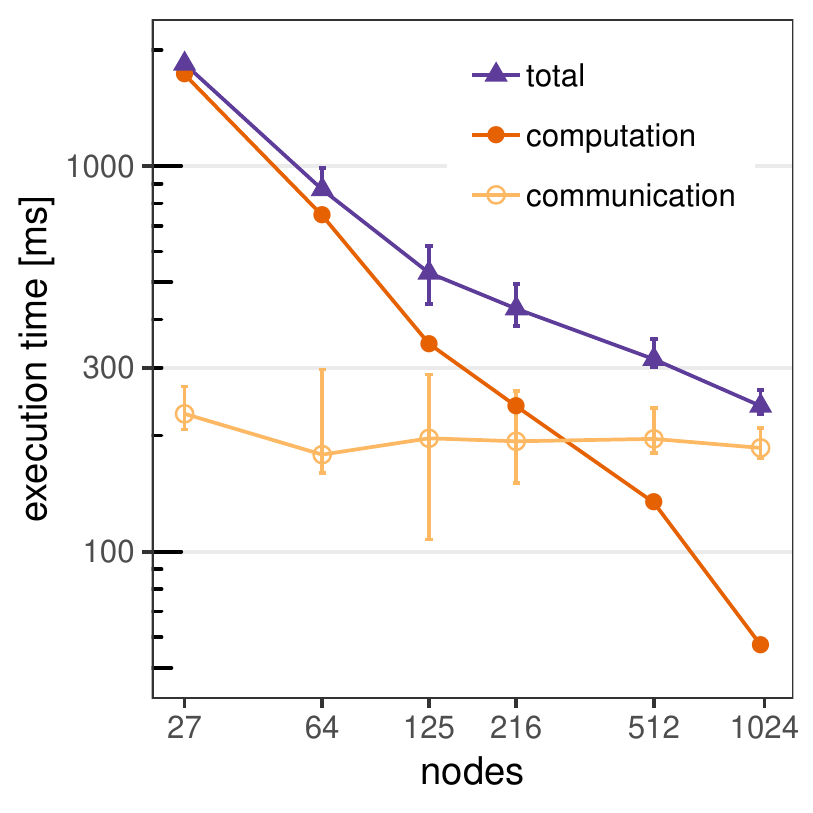}
      \includegraphics[width=0.45\linewidth]{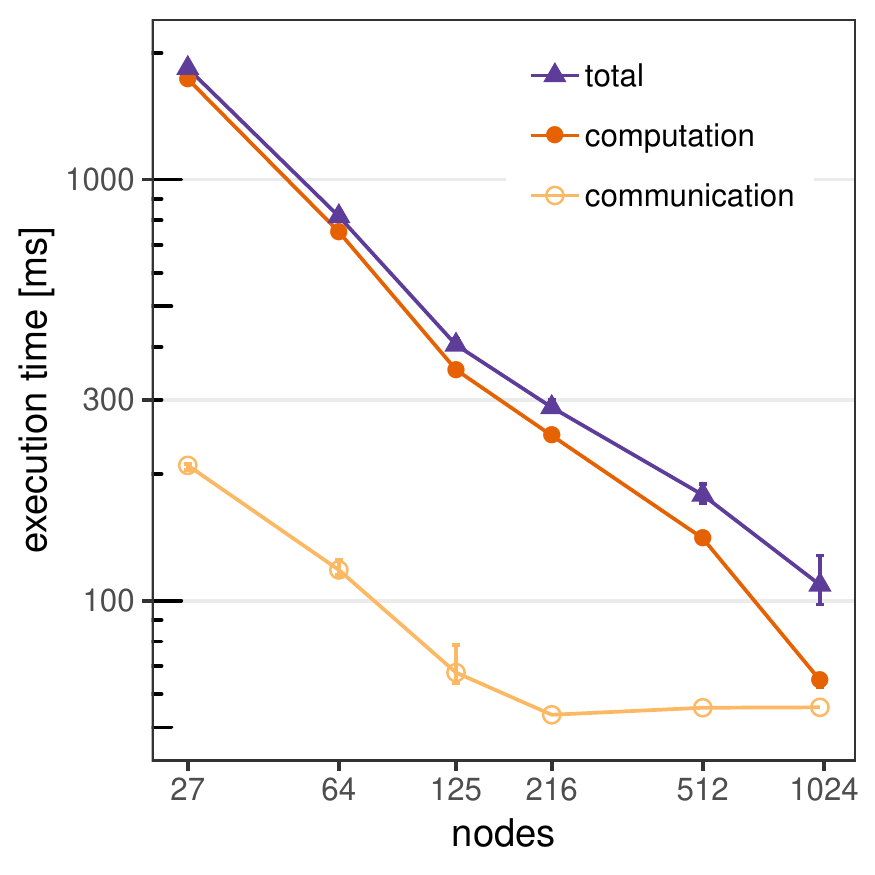}
      \put(-335,150){\bf (a)}
      \put(-165,150){\bf (b)}
      \caption{Scaling of computation, communication and total simulation times for (a) TS algorithm and (b) WS algorithm using notified access communication model in a strong scaling exercise with a problem size of $\ngp{120}$.}
      \label{fig:balance120}
\end{figure}

To study the performance and scalability, we evaluate the weakly-synchronous
algorithm (WS) for different process/node counts and domain sizes, and compare its execution times to the traditional synchronous (TS) algorithm. Two sets of experiments examine strong and weak scaling for configurations relevant on future exascale systems and for setups that are commonly used today. The configurations scale the sub-domain size down to $\ngp{12}$ and $\ngp{24}$ grid points per process and compute on up to $\ngp{120}$ and $\ngp{240}$ grid points in total. The smallest configuration assigns $12^2$ grid points to every thread. As shown in \autoref{fig:balance120}, this extreme scaling has comparable computation and communication times, and therefore fully utilizes the costly interconnect and compute hardware. 

First, we present results from the strong scaling experiments. \autoref{fig:balance120} shows the computation, communication and total simulation times for a problem size of $\ngp{120}$. The runs decompose the domain evenly along all dimensions, resulting in cubical sub-domains. The process count is increased from $27$ to $1000$. 
Part (a) of the figure is obtained from the TS-NA algorithm. As the process 
count increases, the number of grid points per process decreases, and hence the 
computation time steadily reduces. 
Although the communication volume per process also decreases with the process count,
the communication time remains nearly constant due to increasing synchronization 
cost and overwhelms the computation time above $250$ processes. 
A significant contribution to the total 
time is observed starting from $125$ processes, resulting in a poor 
scalability. The error bars in the graph indicate a substantial variability in 
communication time due to system noise among different runs, particularly at 
intermediate process counts. This variability also propagates into the total 
time, however to a lesser extent. Results from the WS algorithm are shown in 
\autoref{fig:balance120} (b). As the computational part of the code remains the 
same, the measured computation time remains unchanged compared to the TS 
algorithm. On the other hand, the communication time is significantly lower for 
the WS algorithm, with a speedup of 3.3x at the extreme scale. This is due to 
the relaxed synchronization that also dampens the performance variability. 
As a result, the bandwidth rather than the latency of the interconnect 
limits the communication performance of the WS algorithm.
%
As the communication time does not exceed the computation time, a significant improvement in the scaling of the total time is observed.  

\begin{figure}
      \centering
      \includegraphics[width=0.45\linewidth]{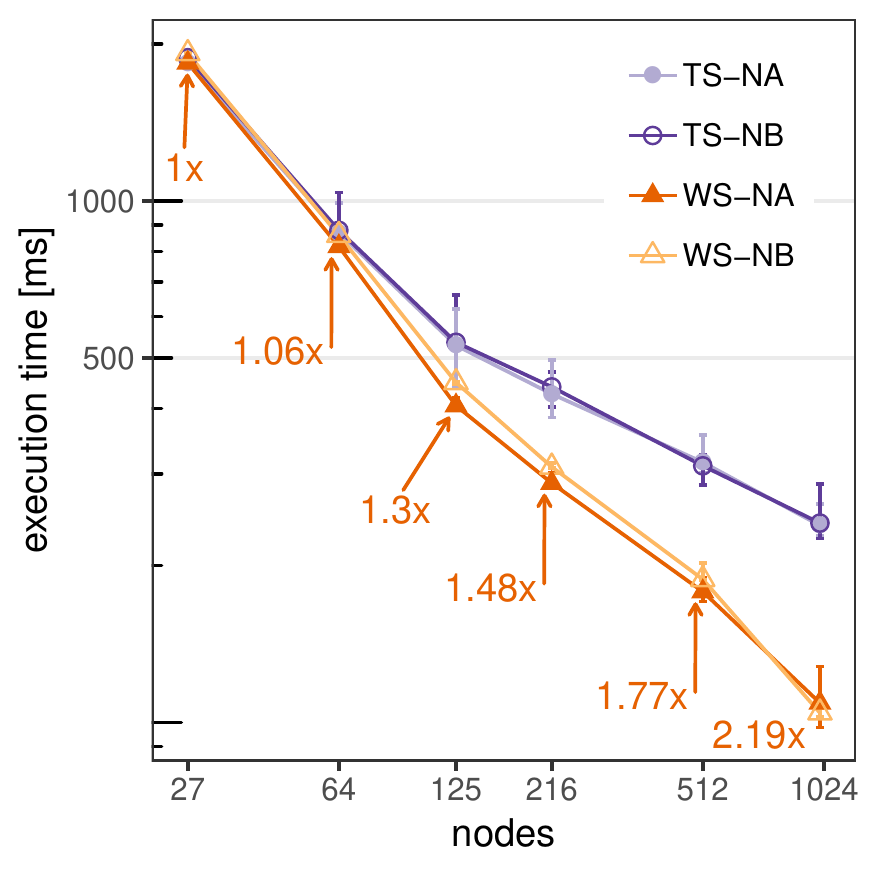}
      \caption{Strong scaling of $\ngp{120}$ problem size. TS: traditional-synchronous algorithm, WS: weakly-synchronous algorithm, NA: one-sided notified access communication model, NB: two-sided non-blocking communication model.}
      \label{fig:strong}
\end{figure}

\begin{figure}
      \centering
      \includegraphics[width=0.45\linewidth]{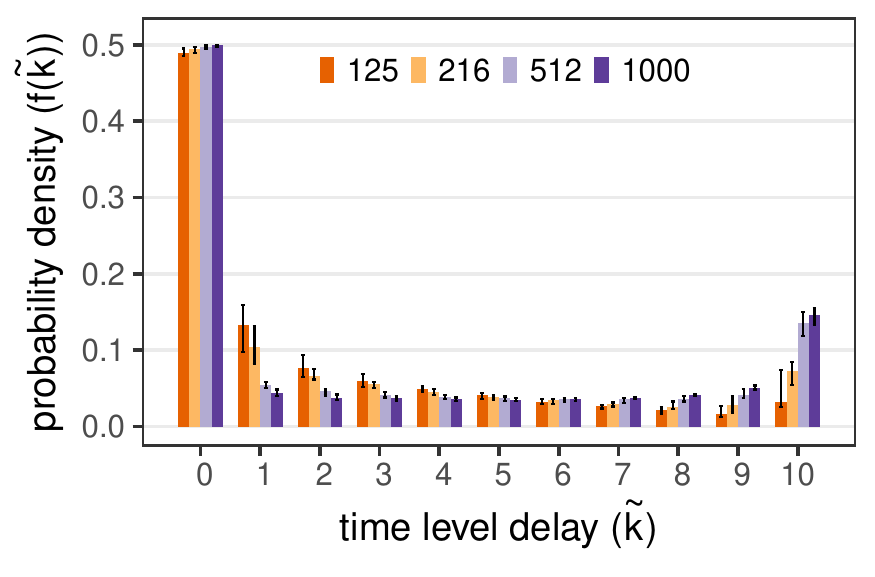} \hspace{0.25cm}
      \includegraphics[width=0.45\linewidth]{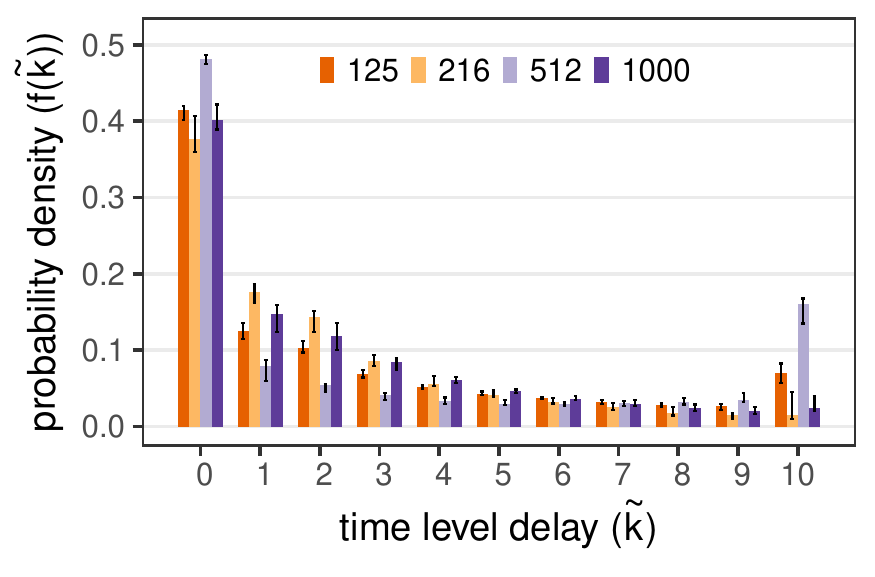}
      \put(-357,85){\bf (a)}
      \put(-177,85){\bf (b)}
      \caption{Distribution of delay in time level ($\k{}$) in WS algorithm simulations for a problem size of $\ngp{120}$ at different node counts. (a) WS algorithm with one-sided notified access communication model, (b) WS algorithm with two-sided non-blocking communication model.}
      \label{fig:delay120}
\end{figure}

\begin{figure}
      \centering
      \includegraphics[width=0.5\linewidth]{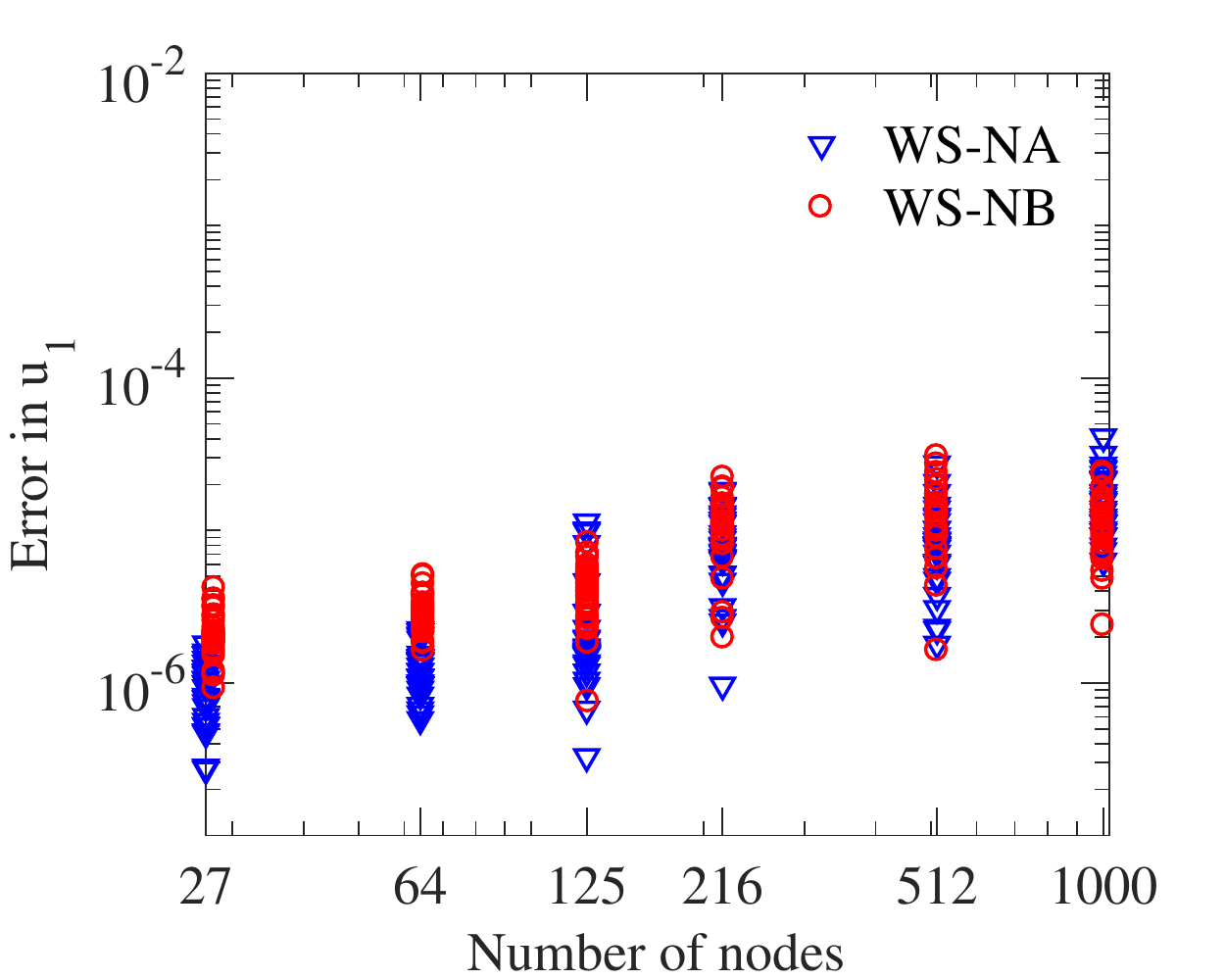}
      \caption{Average error in $u_1$ component of velocity from the weakly-synchronous (WS) algorithm computed against the synchronous solution for different runs in the strong scaling experiments of $\ngp{120}$ grid points problem size. Triangles and circles are from the simulations with one-sided notified access (NA) and two-sided non-blocking (NB) communication models, respectively. }
      \label{fig:error120}
\end{figure}

The overall strong scaling of the two algorithms in combination with the two communication models for $\ngp{120}$ grid problem size is shown in \autoref{fig:strong}. Clearly, the WS algorithm outperforms the TS algorithm with a maximum speed up of $2.19$x at the extreme scale. The one-sided notified access communication model provides an additional minor improvement in the scaling. For the WS algorithm, the distribution of time delay in the function value experienced at the buffer points ($\k{}$) for the two communication models is presented in \autoref{fig:delay120}. As discussed earlier, moments of the distribution of $\k{}$ effect the accuracy of simulations. The figure shows that a better distribution with lower mean and variance is obtained with the NA communication model. Also the spread in the distributions across different runs is also lower for the NA communication model. To further assess the accuracy of solution, we compute the error in each of the WS algorithm runs by comparing its solution with the synchronous solution. \autoref{fig:error120} shows the average error in the $u_1$ component of velocity for the runs at different node counts. It can be inferred that the NA communication model, due to a better distribution of the delays, provides relatively lower error values. 

\begin{figure}
      \centering
      \includegraphics[width=0.45\linewidth]{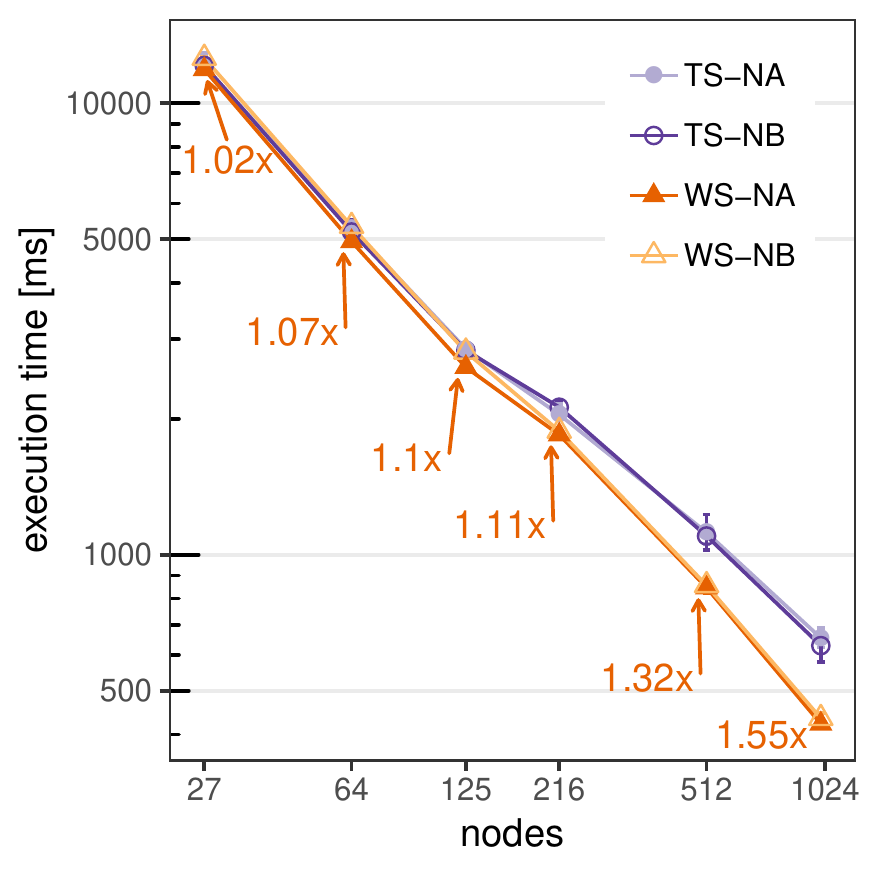}
      \caption{Strong scaling of $\ngp{240}$ problem size. TS: traditional-synchronous algorithm, WS: weakly-synchronous algorithm, NA: one-sided notified access communication model, NB: two-sided non-blocking communication model.}
      \label{fig:strong240}
\end{figure}

The strong scaling of a larger problem size with a grid of $\ngp{240}$ is illustrated in \autoref{fig:strong240}. The WS algorithm continues to outperform the TS algorithm. However, the speed up at the extreme scale decreases to $1.55$x. This can be attributed to the increase in computational effort per process, and nearly the communication. A marginal gain is observed due to the NA communication model. 

\begin{figure}
      \centering
      \includegraphics[width=0.45\linewidth]{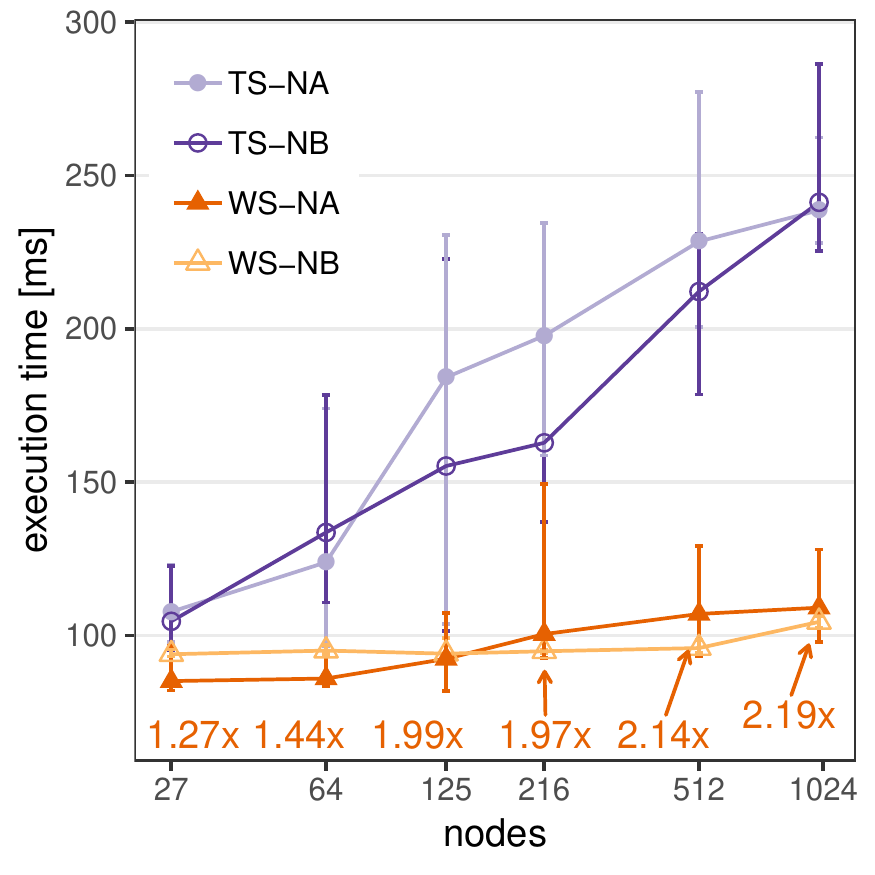}
      \includegraphics[width=0.45\linewidth]{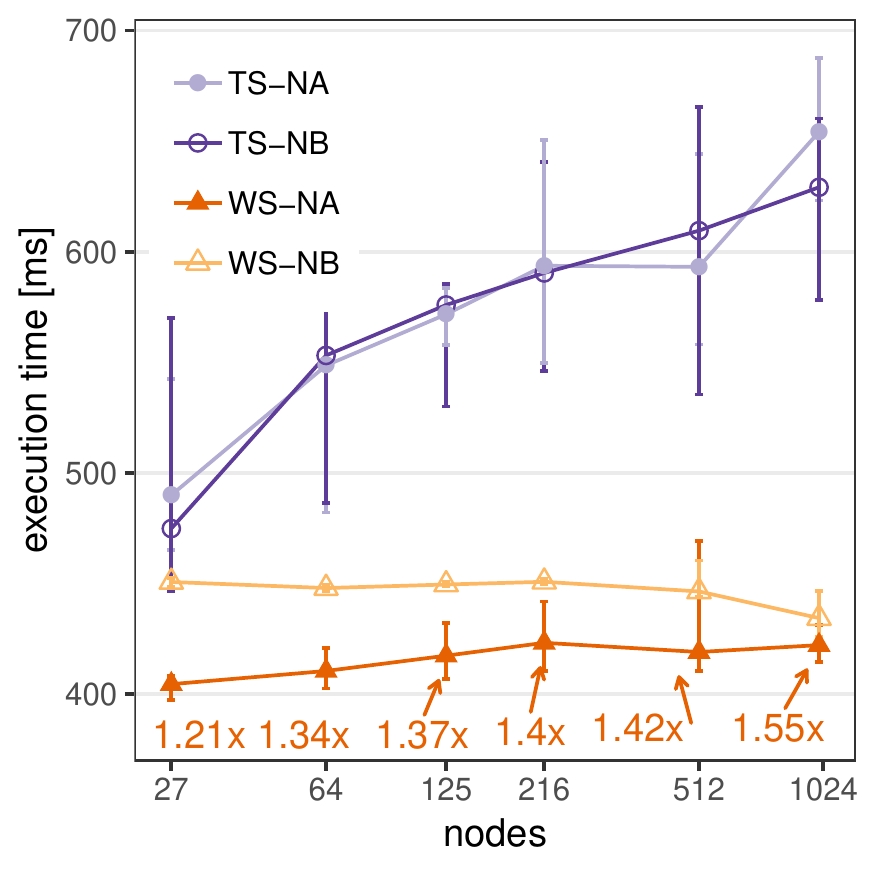}
      \put(-337,145){\bf (a)}
      \put(-166,145){\bf (b)}
      \caption{Weak scaling of (a) $\ngp{12}$ and (b) $\ngp{24}$ grid points per process. TS: traditional-synchronous algorithm, WS: weakly-synchronous algorithm, NA: one-sided notified access communication model, NB: two-sided non-blocking communication model.}
      \label{fig:weak}
\end{figure}

The results from the weak scaling for $\ngp{12}$ and $\ngp{24}$ grid points per process are shown in \autoref{fig:weak}. The runs again decompose the domain evenly along all dimensions, which expands the overall compute domain from $\ngp{36}$ to $\ngp{120}$ and from $\ngp{72}$ to $\ngp{240}$ for the two configurations. The execution time for the WS algorithm remains nearly constant, while it significantly increases for the TS algorithm. This shows that WS algorithm is able to fully overlap the communications with computations. A speed up from $1.27$x to $2.19$x and $1.21$x to $1.55$x  are observed in the $\ngp{12}$ and $\ngp{24}$ grid points per process configurations, respectively. The weak scaling also shows that the WS algorithm possess a lesser spread in the execution times among different runs. This demonstrates the robustness of the algorithm to system noise, consistent with our finding in the strong scaling.

\section{Related Work}\label{sec:related}
This work describes an asynchronous computing approach for solving PDEs, which
uses an explicit time discretization for time derivatives. This relieves the need to solve a system of linear algebraic equations.
However, certain classes of problems are governed by PDEs that are stiff in nature.
Such PDEs are commonly solved with implicit time integration schemes, which result in linear system of equations that have to be solved at each time advancement.

Asynchronous iterative solvers, which do not require synchronization at
each iteration, are available in the literature for linear systems \cite{BT89,FS00}. In
the context of solving PDEs, these methods are useful only when
time-implicit schemes are used. Even in such cases computations of linear
systems within a time step can proceed asynchronously, but it is still
necessary to communicate the data and synchronize at the end of each
step.
This issue can be resolved by coupling the asynchronous solvers with the 
 the weakly-synchronous algorithm for PDEs.

Asynchronous methods for PDEs have been earlier proposed in Amitai at 
al.~\cite{AAII1996,AAII1998}, but their treatment is restricted to parabolic 
PDEs. They use corrections based on Green's function to account for the 
asynchrony at process boundaries. In more recent works 
\cite{Mudigere2014,mittal2016}, relieving the affects of  asynchrony in 
communication have been addressed by modifying the governing PDEs instead of 
the numerical method. They demonstrated mathematical feasibility of their 
method using simple PDEs, but did not evaluate the computational performance. 
Also, extending their work to complex equations seems to be difficult. 
Methods that address other issues like resiliency to system faults and noise have been proposed in \cite{sargsyan2015,grout2015}.

Multiple works discuss the technology aspects of implementing efficient halo 
exchange communication. Kjolstad et al.~\cite{ghost} avoid synchronization 
using the non-blocking interface of two-sided MPI. Gropp et al.~\cite{rmaperf} 
demonstrate the benefits of one-sided MPI on platforms with native hardware 
support. Szpindler~\cite{fompihalo} compares the performance of different 
one-sided and two-sided implementations and obtains promising results for the 
foMPI-NA based version.

\section{Conclusions}\label{sec:conclusions}
At extreme scales, data communication and synchronization between PEs significantly affect the scalability of parallel solvers.
On future exascale machines which would be comprised of millions of PEs, these 
issues 
will certainly amplify further, and are likely to pose a bottleneck in 
designing scalable solvers.
In an effort to relax synchronization and minimize communication costs in PDE 
solvers, we proposed a novel weakly-synchronous (WS) algorithm based on a 
mathematically asynchronous numerical method, where computations at processes 
can advance regardless of the status of communications. 
The numerical method uses new asynchrony-tolerant (AT) schemes that maintain 
the accuracy of the computed solution in presence of data asynchrony.
However, these schemes pose three algorithmic requirements, namely: (a) 
synchronize communications between processes only when the delay between them 
is greater than the maximum allowable delay, (b) storage of data from multiple 
time levels or iterations at buffer points, and (c) knowledge of the time 
level 
associated with each communication.
The WS algorithm addresses these requirements with minimal computation and 
communication overhead. The algorithm uses an efficient notified remote memory 
access programming model to implement communication between processes.
The relaxed synchronization between processes naturally leads to a 
computation-communication overlap and has the potential to relieve the effects 
of system noise.

The performance of the WS algorithm is shown in comparison with traditional synchronous (TS) algorithm. We have developed a mini-application that simulates the multi-scale Burgers' turbulence phenomena for the benchmark experiments. We use numerical schemes that provide an overall second order accuracy of solution in space.
The algorithm implementation would remain similar for higher order accurate methods. 
The results from the numerical experiments show that the accuracy of the solution in WS 
algorithm are marginally affected due to data asynchrony. The WS algorithm 
provides a significant speed up over the synchronous algorithm. A speed up of 
2.19x has been observed at extreme scales. The WS algorithm is also shown to be more robust to system noise.
The performance evaluation of the WS algorithm clearly exhibits the potential for designing highly scalable PDE solvers for future exascale machines.

\section*{Acknowledgments}
We thank the Swiss National Supercomputing Center for providing the computing resources and for supporting us with the benchmark setup.
The work at Sandia National Laboratories was supported by the US Department of Energy, Office of Basic Energy Sciences, Division of Chemical Sciences, Geosciences, and Biosciences. Sandia National Laboratories is a multimission laboratory managed and operated by National Technology and Engineering Solutions of Sandia, LLC., a wholly owned subsidiary of Honeywell International, Inc., for the US Department of Energy's National Nuclear Security Administration under contract DE-NA-0003525. The views expressed in the article do not necessarily represent the views of the U.S. Department of Energy or the United States Government.


%


\bibliographystyle{model1-num-names}
\bibliography{main.bib}




%



\end{document}